%Paper: gr-qc/9310026
%From: thooft@ruuntk.fys.ruu.nl (Gerard 't Hooft)
%Date: Tue, 19 Oct 93 14:17:10 GMT

% Figures in files attached.

\magnification= \magstep1
\tolerance=1600
\parskip=0pt
\baselineskip= 6 true mm
 \input epsf
\font\smallrm=cmr8

\def\a{\alpha}
\def\b{\beta}
 
\def\d{\delta}

\def\s{\sigma}

\def\cl{\centerline}
\def\ts{\textstyle}
\def\x{{\bf x}}
\def\E{{\bf e}}
{\nopagenumbers

\vglue 1truecm \rightline{THU-93/26} \rightline{gr-qc/9310026} \vskip 2 truecm

\cl {\bf DIMENSIONAL REDUCTION in QUANTUM
GRAVITY\footnote{$\,^\dagger$}{\smallrm Essay dedicated to Abdus Salam}}
\vskip 1 truecm
\cl {G. 't Hooft}
\cl {Institute for Theoretical Physics}
\cl {Utrecht University}
\cl {Postbox 80 006, 3508 TA Utrecht, the Netherlands}
\vfil
{\bf Abstract}
\medskip
The requirement that physical phenomena associated with gravitational
collapse should be duly reconciled with the postulates of quantum
mechanics implies that at a Planckian scale our world is not 3+1
dimensional. Rather, the observable degrees of freedom can best be
described as if they were Boolean variables defined on a
two-dimensional lattice, evolving with time. This observation, deduced
from not much more than unitarity, entropy and counting arguments,
implies severe restrictions on possible models of quantum gravity.
Using cellular automata as an example it is argued that this
dimensional reduction implies more constraints than the freedom we have
in constructing models. This is the main reason why so-far no
completely consistent mathematical models of quantum black holes have
been found.

\vfil\eject}
With the request to write a short paper in honor of Abdus Salam I am
given the opportunity to contemplate some very deep questions
concerning the ultimate unification that may perhaps be achieved when
all aspects of quantum theory, particle theory and general relativity
are combined. One of these questions is the dimensionality of space and
time.

At first sight our world has three spacelike dimensions and one
timelike. This was used a a starting point of all quantum field
theories and indeed also all string theories as soon as they invoke the
Kaluza Klein mechanism. Also when we quantize gravity perturbatively we
start by postulating a Fock space in which basically free particles
roam in a three plus one dimensional world.  Naturally, when people
discuss possible cut-off mechanisms, they think of some sort of lattice
scheme either in 3+1 dimenisional Minkowski space or in 4 dimensional
Euclidean space. The cut-off distance scale is then suspected to be the
Planck scale.

Unfortunately any such lattice scheme seems to be in conflict with
local Lorentz invariance or Euclidean invariance, as the case may be,
and most of all also with coordinate reparametrization invariance. It
seems to be virtually impossible to recover these symmetries at large
distance scales, where we want them. So the details of the cut-off are
kept necessarily vague.

The most direct and obvious {\it physical} cut-off does not come from
non-renormalizability alone, but from the formation of microscopic
black holes as soon as too much energy would be accumulated into too
small a region. From a physical point of view it is the black holes
that should provide for a natural cut-off all by themselves.

This has been this author's main subject of research for over a decade.
A mathematically consistent formulation of the black hole cut-off turns
out to be extremely difficult to find, and in this short note I will
explain what may well be the main reason for this difficulty: nature is
much more crazy at the Planck scale than even string theorists could
have imagined.

One of my starting points has been that quantum mechanics itself is not
at all a mystery to me. The emergence of a Hilbert space with a
Copenhagen interpretation of its inner products is a quite natural
feature of any theory with the following characteristics at a local
scale: the system must have {\it discrete} degrees of freedom at tiny
ditance scales, and the laws of evolution must be {\it reversible} in
time. With discrete degrees of freedom one can construct Hilbert space
in a quite natural way by postulating that any state of the physical
degrees of freedom corresponds to an element of a basis of this Hilbert
space[1]. Reversibility in time is required if we wish to see a quantum
superposition principle: the norm of all states is then preserved, even
if they are quantum superpositions of these basis elements.

Needless to say, one might suspect that some or all of the quantum
mechanical postulates could break down at the Planck scale[2]. But then
one might as well throw away anything we know about physics, and that
is not the route I want to follow. I have never seen convincing models
where ordinary quantum mechanics breaks down at a microscopical level
but is somehow recovered at the atomic scale. Therefore I prefer not to
speculate that quantum mechanics breaks down at the Planck scale, but
in stead to suspect that quantum mechanics becomes trivial there:
quantum superpositions are still allowed there but become irrelevant.

Black holes then present a major challenge. At first sight they render
time reversibility impossible. Objects thrown into a black hole can
never be retrieved[3]. Things are often presented as if black holes
connect our world to other universes via wormholes[4], or, if  one
prefers not to refer to these other universes one says that information
thrown in can not be retreived anymore[5]. According to this view
information is preserved only if one considers multiply connected
universes[6]. This is useless if one wishes to recover quantum
mechanical behavior in our universe by itself. However, at closer
inspection one finds that any object thrown into a black hole actually
does leave some signals behind in our own world[7], and it is
conceivable that a unitary theory for our own universe can be built
using this as a starting point.[8]

The main difficulty then is to formulate exactly what our derees of
feedom are. Remarkably, it is relatively easy to give a fairly precise
estimate of {\it how many} degrees of freedom we have. This can be
deduced in two equivalent ways[9, 10]. One is by considering capture and
emission of objects by black holes as scattering experiments. If these
are described quantum mechanically one can deduce information about the
size of phase space. It must be finite, increasing exponentially with
the surface area of the black hole horizon. The other way to deduce the
same information is by using thermodynamics. One derives the entropy
$S$ of a black hole, finding
$$S=4\pi M^2+C \,,\eqno(1)$$
in natural units. The constant $C$ is not known (in fact there could be
as yet unknown subdominant terms in this expresion, increasing slower
than $M^2$ as the mass increases). In principle $C$ could be infinite
(even for small $M$), which basically corresponds to the remnant
theory. I think an infinite $C$ would induce major problems (again
implying a deviation from ordinary quantum mechanical behavior) in a
consistent theory, so I will henceforth assume $C$ to be bounded.

This entropy is often attributed `to the space-time metric itself', as
if there would be yet another separate contribution from the quantized
fields in this metric, such as the contribution to the entropy of
objects outside the black hole. However, this would not be correct.
First of all, the contribution of objects at some distance from the
black hole will always be negligible unless they are really very far
away (at $|x|\gg R_{\rm Schwarzschild}$). But most importantly, the
contribution of quantum fields to the entropy {\it diverges} very near
the horizon[10]. The reason for this divergence can be easily understood
physically: arbitrary amounts of matter can be thrown in and arbitrary
amounts are waiting to radiate out; they contribute infinitely to the
total number of physical degrees of freedom.

Yet the black hole entropy had just been argued to be finite, {\it even
with the quantum fields present!} We conclude that one has to attribute
the black hole entropy not to the space-time metric itself but to the
quantized fields present there (which of course does include the small
quantum fluctuations of the metric itself), and then one must choose a
cut-off sufficiently close to the horizon such that it exactly matches
the known black hole entropy (1). In short, the black hole entropy {\it
includes} the entropy of the quantized fields in its neighborhood[10].

In any quantum theory there is a `third law of thermodynamics' relating
the entropy to the total number of degrees of freedom: the dimension of
the vector space describing all possible states our system can be in is
the exponent of the entropy. For instance in a discrete theory
described by $n$ spins that can take only two values (`Boolean
variables'), the dimension $\cal N$ of Hilbert space is
$$e^S={\cal N}=2^n\ ,\eqno(2)$$
Hence the entropy directly counts the number of Boolean degrees of freedom.

Considering the fact that at the Hawking temperature the contribution
to the entropy of fields anywhere in the region $R<|x|\ll R^3$ pales
compared to the black hole entropy itself, we can now make an important
observation concerning the relevant degrees of freedom of a black
hole:  \smallskip {\it The total number of Boolean degrees of freedom,
$n$, in a region of space-time surrounding a black hole is
$$n={S\over\ln 2}={4\pi M^2\over\ln 2}={A\over 4\ln2}\ ,\eqno(3)$$
where $A$ is the horizon area.}
\smallskip

We can carry this argument one step further. Consider just any closed
spacelike surface, with $S(2)$ topology and total surface area $A$.
Consider all possible field and metric configurations inside this
surface. We ask how many mutually orthogonal states there can be. If we
want these states to be observable for the outside world we have to
insist that the total energy inside the surface be less than $1/4$
times its linear dimensions, otherwise our surface would lie within the
Schwarzschild radius. Let us first ask how many states would an
ordinary quantum field theory allow us to have, given these limits on
the volume $V$ and the energy $E$. Now this is not hard. The most
probable state would be a gas at some temperature $T=1/\b$. Its energy
would be approximately
$$E=C_1Z V T^4\,,\eqno(4)$$
where $Z$ is the number of different fundamental particle types with mass less
than $T$ and $C_1$ a numerical constant of order one, all in natural units.
The total entropy $S$ is
$$S=C_2Z V T^3\,,\eqno(5)$$
where $C_2$ is another dimensionless constant. Now the Schwarzschild limit
requires that
$$2E<(V/ \ts{4\over3}\pi)^{1\over3}\,,\eqno(6)$$
hence, with eq. (4),
$$T<C_3Z^{-{1\over4}}V^{-{1\over6}}\,,\eqno(7)$$
so that
$$S<C_4Z^{1\over4}V^{1\over2}=C_5Z^{1\over4}A^{3\over4}\,.\eqno(8)$$
The $C_i$ are all constants of order 1 in natural units.  Since in
quantum field theories, at sufficiently low temperatures, $Z$ is
limited by a dimensionless number we find that this entropy is small
compared to that of a black hole, if the area $A$ is sufficiently
large.

Next consider a set of $N$ black holes, with masses $M_i$. They
contribute to the energy $\sum_i M_i$. So $$\sum_i
M_i<C_6A^{1\over2}\,,\eqno(9)$$ while their total entropy is given by
$$S=C_7\sum_iM_i^2\,,\eqno(10)$$ where we note that the contribution of
their movements to the entropy is negligible. We see that ineq. (10) is
saturated when one black hole has the largest possible size that still
fits inside our area. Its entropy is $$S_{max}=
\ts{1\over4}A\,,\eqno(11)$$ and this is as large as we can ever make
it. This therefore answers our question concerning the total number of
possible states. It is given by eqs. (2) and (3). The single black hole is the
limit[11].

The importance of this result can hardly be overestimated. At first
sight it is counterintuitive. One would have expected that the number
of possible states would grow exponentially with the volume, not the
area, as in any ordinary field theory with a cut-off. So if one would
take a regularized fermion theory with cut-off at the Planck scale one
would get far too many states. But it is clear why our answer came out
this way. Most of the states of a regularized quantum field theory
would have so much energy that they would collapse into a black hole
before they could dictate the further evolution of the system in time.
If we want to avoid this a much more rigorous cut-off than a Planckian
one must be called for. Note that if there are any fundamental bosons
the number of possible states comes out to be strictly infinite.

But then one may come to appreciate this result after all. It means
that, given any closed surface, we can represent all that happens
inside it by degrees of freedom on this surface itself. This, one may
argue, suggests that quantum gravity should be described entirely by a
{\it topological} quantum field theory, in which all physical degrees
of freedom can be projected onto the boundary. One Boolean
variable per Planckian surface element should suffice. The fact that
the total volume inside is irrelevant may be seen as a blessing since
it implies that we do not have to worry about the {\it metric} inside.
The inside metric could be so much curved that an entire universe could
be squeezed inside our closed surface, regardless how small it is. Now
we see that this possibility will not add to the number of allowed
states at all. Our result suggests that one should not worry about
creating universes inside test tubes.

The same can be said about wormholes. A wormhole with one end sprouting
inside our closed volume and its other end somewhere else could connect
the inside of our volume to the outside world, thus adding large
quantities of possible states. We now believe that this is not allowed
in a decent theory of quantum gravity. In a previous publication [8] it
was explained why the functonal integral describing black holes
probably has to be limited to topologically trivial field
configurations only. We have detailed ideas concerning the consistency
of such a requirement but will not elaborate on this here. So much for
wormholes.

We would like to advocate here a somewhat extreme point of view. We
suspect that there simply {\it are} not more degrees of freedom to talk
about than the ones one can draw on a surface, as given by eq. (3) The
situation can be compared with a hologram of a three dimensional image
on a two-dimensional surface. The image is somewhat blurred because of
limitations of the hologram technique, but the blurring is small
compared to the uncertainties produced by the usual quantum mechanical
fluctuations. The details of the hologram on the surface itself are
intricate and contain as much information as is allowed by the
finiteness of the wavelength of light - read the Planck length.

It is tempting to take the limit where the surface area goes to
infinity, and the surface is locally approximately flat. Our variables on
the surface then apparently determine all physical events at one side (the
black hole side) of the surface. But since the entropy of a black hole
also refers to all physical fields outside the horizon the {\it same}
degrees of freedom determine what happens at this side. Apparently one
must conclude that a two-dimensional surface drawn in a three-space can
contain all information concerning the entire three-space. In fact,
this should hold for {\it any} two-surface that ranges to infinity.
This suggests that physical degrees of freedom in three-space are not
independent but, if considered at Planckian scale, they must be
infinitely correlated.

If one could determine the equation of motion of these variables on the
two-plane one would also possess the remedy for the black hole
information paradox. In a Rindler space we could put our information
surface at the origin of the Rindler coordinate frame. The
transformation rules for our variables under Lorentz transformations
would correspond to the Rindler equations of motion, and pure states
would evolve into pure states while the spectrum density of the black
hole would continue to be the one dictated by its entropy.

The infinite correlations in three-space could also present a new
starting point for resolving the Einstein-Rosen-Podolsky paradox. In
terms of present day quantum field theoretical degrees of freedom it is
not possible to interpret quantum mechanics deterministically. But
certain cellular automaton models can give a faint resemblace to
quantum behavior. The discreteness of our degrees of freedom on the
two-surface strongly remind us of a 2+1 dimensional cellular automaton.
But this would constitute a speculation on top of the previous one.
There are various technical problems one has to face if such
considerations were to be persecuted further.

In cellular automaton models for quantum mechanics the problem is not
the Copenhagen interpretation[1]. This comes out quite naturally. The
more tantalizing problem is how to understand the stability of our
vacuum state[13].  Models can be constructed producing Hamiltonians that
can be realistic but for which there are nearly always negative energy
eigenstates. The absence of negative energy states in the real world is
probably related to the presence of the gravitational force, for which
the time coordinate is handled very differently from non-gravitational
theories. But natural curvature of space and time is equally difficult
to realize in cellular automaton models of this sort[12]; we just suspect
that these problems are related but exactly how is not understood.

But there are other problems as well. One of these is the presence of
the group of Lorentz transformations and the fact that this symmetry
group is non-compact. This is of course at the heart of the black hole
horizon problem.  There, Lorentz transformations are playing the role
of time translations. In any theory where the physical degrees of
freedom are discrete it is extremely difficult to reproduce anything
resembling Lorentz invariance \footnote{$\, ^1$}{\smallrm In principle
one could think of realizing one of the non-trivial discrete subgroups
of the Lorentz group, but in practice this seems to be impossible to
reconcile with locality requirements.}. As for other invariances such
as rotational invariance, one can usually realize symmetry under one of
their finite subgroups and then it is not unnatural to suspect that
complete invariance will be recovered in the thermodynamic limit as a
consequence of renormalization group effects.

The evolution law of the physical degrees of freedom on a 2-surface is
another mystery. Ultimately one wishes to recover full invariance with
respect to the Poincar\'e group (which is a precisely defined
invariance group for all states that have asymptotic in- and
out-states, as the ones used in an $S$ matrix formalism, but not in
quantum cosmology). Now this implies that one not only needs an
evolution law for translations in the time direction but also a quite
similar looking law for translations in the direction orthogonal to the
surface. These two evolution laws should commute with each other in
spite of their complete independence. We will show how to reproduce
such a feature in cellular automaton models, however, as we will see,
requiring these models to possess physically interesting, that is,
sufficiently non-trivial interactions, will remain a difficult
obstacle.

The question whether models exist in 3+1 dimensions that are such that
the data on a two-dimensional surface will determine all observables
elsewhere is an extrelely intriguing one. For definiteness, let us
consider a rectangular lattice in 4-space. Momentarily we will ignore
requirements such as Lorentz covariance; it is sufficient to require
that signals do not go faster than some limited speed $c$. The
prototype of our models is a cellular automaton. The data $f$ on
every site on the lattice can be represented by an integer modulo some
number $p$. This $p$ will often be taken to be a prime number. The time
evolution is defined by some local law. Conventionally, one imposes
that the value of $f(\x,t)$ be a given function of the values of
$f(\x_i,t-1)$, where $\{\x_i\}$ are a finite set of nearest neighbors
of $\x$.

Now such a model is deterministic in the classical sense. Quantization
can be introduced either by allowing $f(\x_i,t)$ to be operators in
Hilbert space, or even more simply, by declaring all states of the
cellular automaton to be basis elements of Hilbert space.
"Quantization" is then trivial. There are problems with this latter
proposal in that the Hamiltonian then does not seem to possess a
well-defined ground state. Again, let's not dwell on that.

In order to obtain "dimensional reduction" we will have to postulate a
further constraint: the values of $f$ on a sheet should fix the values
elsewhere.  This we can realize in principle by also postulating a law
of evolution in the $z$ direction. Since further discussion of this 4
dimensional problem becomes a bit intricate it is illustrative to
remove one dimension and treat space as 2 dimensional, space-time as 3
dimensional. Here we will require that the data on a {\it line} should
be sufficient to determine the data elsewhere. Part of the rectangular
lattice is depicted in Fig. 1.

Suppose now that on every plaquette of the lattice a relation among the
data $f$ is imposed. Thus, in the figure there are constraints of the
form
$$ \eqalignno {
g_a\big(f(A),f(B)&,f(C),f(D)\big)=0\,,&(12a)\cr
g_b\big(f(E),f(F)&,f(G),f(H)\big)=0\,,&(12b)\cr
g_c\big(f(A),f(B)&,f(F),f(E)\big)=0\,,&(12c)\cr
g_d\big(f(D),f(C)&,f(G),f(H)\big)=0\,,&(12d)\cr
g_e\big(f(A),f(D)&,f(H),f(E)\big)=0\,,&(12e)\cr
g_f\big(f(B),f(C)&,f(G),f(F)\big)=0\,.&(12f)\cr}$$
We require all these constraints to be such that if in any of these
equations three of the four entries are given the fourth will be
uniquely determined.

\midinsert \vskip 6 truecm \epsffile {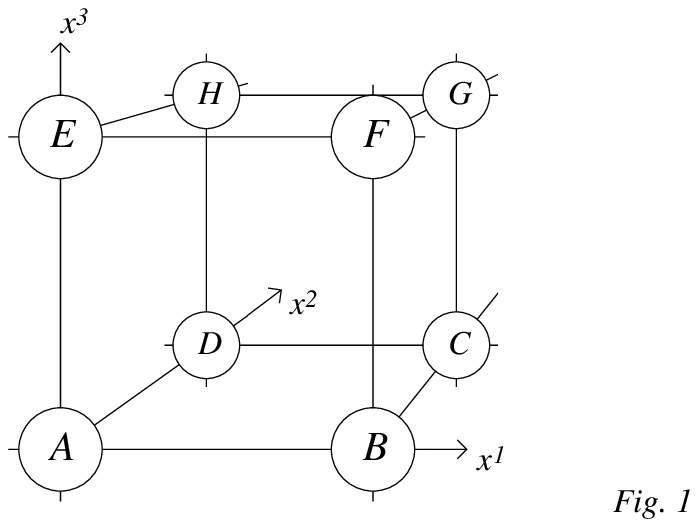} % \special{thfig1.ps}

\narrower\narrower{\smallrm The lattice sites in $x^1,x^2,x^3$ coordinates} \endinsert

If we choose the $x,y,t$ coordinates in one of the principle directions
of the lattice the situation becomes a bit singular. It is preferrable
to define $$ \eqalignno {t&=x^1+x^2+x^3\,,\cr
 x&=x^1-x^2\,,&(13)\cr y&=x^3-\textstyle{1\over2}(x^1+x^2)\,.\cr}$$ if
on two nonsecutive $t$ layers the data are known successive application
of eqs (12a--f) will produce the data in all of space-time. But now
consider the series of points $\x(n)$, $n\in Z\!\!\!Z$, defined by $$
\eqalignno{\x(3n+1)&=\x(3n)+\E^1\,,\cr
\x(3n+2)&=\x(3n+1)-\E^2\,,&(14)\cr \x(3n+3)&=\x(3n+2)+\E^3\,,\cr}$$
where $\E^{1,2,3}$ are the three unit vectors of the lattice (the signs
in front of the $\E^i$ can actually be chosen at will). Suppose
$f(\x(n))$ are given for all $n$. Successive application of the six
identities (12 a--f) then also fixes all data elsewhere.

However, one cannot choose the constraints (12a--f) any way one likes.
This is because the data elsewhere will be {\it over-}determined.
Suppose that in the Figure the data are given at $D$, $H$, $E$ and $F$.
They form part of a series (14). By applying four of the six equations
(12) the other data on the cube are determined. The remaining two must
then be satisfied automatically. If these would not be satisfied our
system would be constrained further, in such a way that practically no
solutions survive at all. The point is that eqs (12a, b) generate
translations along the $x^1x^2$ plane, eqs. (12c, d) translations along
the $x^1x^3$ plane and (12e, f) translations along the $x^2x^3$ plane.
These three translation operators, viewed as operators in Hilbert
space, should commute with each other. Only when they are chosen very
meticulously these commutation requirements can be met. One can also
say that we have translation operators defined on the one dimensional
series of data on the points (14). They define translations in the
directions $\E^i\pm \E^j$, which should all commute with each other.
Two linear combinations of these, $U_1(\d t)$ and $U_2(\d y)$ , should
generate {\it independent} translations in directions
orthogonal to the line $\s(\E^1-\E^2+\E^3)$ and a third, $U_3$, corresponds
to a translation in the direction of the line. The three independent
translation operators one then has should commute with each other.
Commutation with $U_3$ is usually easy to implement by choosing the evolution
not to depend on the coordinate $n$, but commutation of $U_1$ with $U_2$
is as hard as finding two different yet commuting local Hamiltonians for
a quantum system.

Let us phrase the constraints the following way. Consider Fig. 1. One is
free to choose $f(A)$, $f(B)$, $f(D)$ and $f(E)$. Then the values of
$f(C)$, $f(F)$ and $f(H)$ are uniquely determined by eqs (12a, c) and (e).
But the value of $f(G)$ is overdetermined. The three equations
(12b, d) and (f) should all yield the same value. Writing the solutions to
eqs (12a--f) as
$$\eqalignno{
f(C)&=h_a\big(f(A),f(B),f(D)\big)\ ,&(15a)\cr
f(G)&=h_b\big(f(E),f(F),f(H)\big)\ ,&(15b)\cr
f(F)&=h_c\big(f(A),f(B),f(E)\big)\ ,&(15c)\cr
f(G)&=h_d\big(f(D),f(C),f(H)\big)\ ,&(15d)\cr
f(H)&=h_e\big(f(A),f(D),f(E)\big)\ ,&(15e)\cr
f(G)&=h_f\big(f(B),f(C),f(F)\big)\ ,&(15f)\cr}$$
our requirement corresponds to \smallbreak
$$\eqalignno{  h_b\Big(f(E),h_c\big(f(A),f(B),f(E)\big),h_e\big(f(A),f(D),
f(E)\big)\Big)&=\cr\nobreak
h_d\Big(f(D),h_c\big(f(A),f(B),f(D)\big),h_e\big(f(A),f(D),f(E)\big)\Big)
&=&(16)\cr
h_f\Big(f(B),h_c\big(f(A),f(B),f(D)\big),h_b\big(f(A),f(B),f(E)\big)\Big)&\,.
\cr}$$

An easy way to implement these commutation constraints is by choosing the
functions $g_i$ to be {\it linear in the functions $f_i$ modulo $p$ }:
$$g_i(\{f_j\})=\sum_j A_{ij}f_j+B_i\ \ {\rm mod}\ p,\eqno(17)$$
where the coefficients $A_{ij}$ must all have an inverse modulo $p$.
{}From this one gets three equations for $f(G)$:  $$f(G)=K_{1,\a}
f(A)+K_{2,\a} f(B)+K_{3,\a} f(D)+K_{4,\a} f(E)+K_{5,\a}\ \ {\rm mod}\ p,\ \
\a=1,2,3.\ \eqno(18)$$
It is not hard to find sets of coefficients $A_{ij},\ B_i$ such that
the 10 equations $$K_{i,1}=K_{i,2}=K_{i,3} \eqno(19)$$ are obeyed.

A next step is to attempt to find other realizations of our
commutation requirement. Using a computer search program the author
generated solutions which at first sight seemed to be quite different,
but careful analysis reveiled that all solutions found were actually
equivalent to the linear ones, eq. (17) after applying permutation
operations on the numbers $f$. This could be seen as a disappointment
because one might argue that a linear relation such as eq. (17) is too
trivial to be of much physical interest. It implies that solutions can
be superimposed onto each other, as if we were describing only
``non-interacting" particles. Our challenge at present is to find any
set of plaquette relations that does not allow a superposition
procedure to obtain new solutions from old ones (note that
superposition here means addition modulo a number, and is quite
distinct from quantum mechanical superposition which is always
allowed).

Our problem could be seen to simplify a bit by using different
lattices.  Basically what we want to achieve is a cellular automaton
with two commuting but essentially different evolution laws. Suppose we
have a triangular lattice in the $x-y$ direction as well as the $x-t$
direction, see Fig. 2. We may demand that the data on the $x$ axis alone
should determine all others. This implies that we have a relation $U_1$
determining how the data look in the $y$ direction, and an evolution
operator $U_2$ in the time direction.  We have $$\eqalignno{
f(F)&=U_1\big(f(A),f(B)\big)\,,&(20a)\cr
f(G)&=U_1\big(f(B),f(C)\big)\,,&(20b)\cr
f(P)&=U_2\big(f(A),f(B)\big)\,,&(20c)\cr
f(Q)&=U_2\big(f(B),f(C)\big)\,,&(20d)\cr
f(X)&=U_2\big(f(F),f(G)\big)\,,&(20e)\cr
f(X)&=U_1\big(f(P),f(Q)\big)\,,&(20f)\cr}$$
and for all choices of $f(A)$, $f(B)$ and $f(C)$ the equations (12e) and
(12f) should agree. These equations are easier to study than the
equations (16). Again there are solutions where $U_1$ and $U_2$ are
different, providing no direct relations between $f(F)$ and $f(P)$ or
$f(G)$ and $f(Q)$, but all solutions we found are equivalent to linear
ones modulo a prime number $p$.

\midinsert \vskip 6 truecm \epsffile {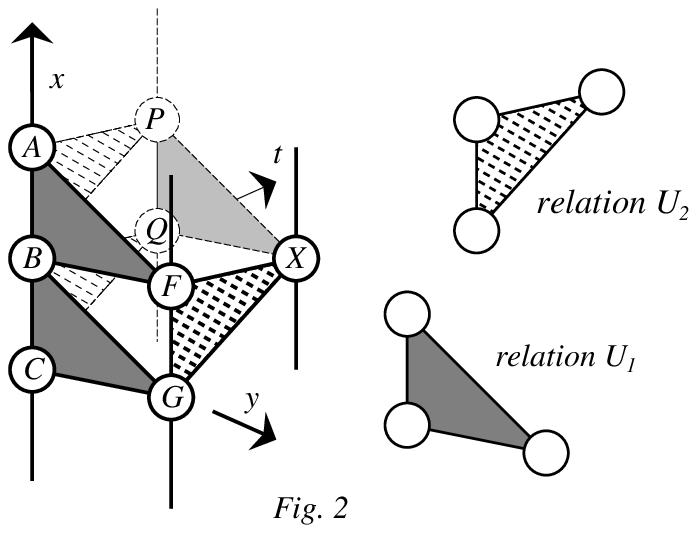}

 \narrower{\smallrm Cellular automaton with two
evolution laws on triangular lattices.} \endinsert

In the real world
there is one more dimension. By choosing the initial data to be
periodic in that one extra dimension the problem reduces to the three
dimensional one. Ultimately one would also like to reobtain local
Lorentz invariance and coordinate reparametrization invariance. These
imply large symmetry groups that are as yet impossible to implement.

The reason why it is interesting to attampt to find an accidental
non-trivial solution of the equations is that many cellular automata.
defined on different lattice types, can be transformed into each other,
as long as the evolution equations act upon nearest neighbors only. So
we actually cover a large class of models. If the real world obeys
comparable rules it could be equaivalent to such an automaton also,
though, admittedly, we consider such a simple structure unlikely.
Indeed, one can also devise interaction schemes among discrete
variables that are essentially local but {\it not} equivalent to
cellular automata of the kind considered here.

Even though the problems just mentioned are grave and conceptually
difficult to disentangle, they do not seem to be insurmountable. Our
basic problem is that there seem to be {\it too many} symmetry
requirements and the Hilbert space of physically realizable states
seems to be {\it too small}. The picture we sketched of the 2+1 (or
perhaps it is better to say 2+2) dimensional nature of our
micro-universe seems to follow from quite general arguments. Rejecting
any of these arguments leads to quite different and perhaps equally if
not more difficult problems, and one cannot help observing that one's
preferences seem to be related to nearly religeous prejudices. Besides
the author not many other physicists have tried this particular avenue.
We advocate its further persuit.
\bigskip
\noindent {\bf References}
\medskip

\item{1.} {G. 't Hooft, {\it Nucl. Phys.} {\bf B 342} (1990) 471;
{\it J. Stat. Phys.}  {\bf 53} (1988) 323}.
\item{2.} S.W. Hawking, {\it Phys. Rev.} {\bf D 14} (1976) 2460.
\item{3.} {S.W.  Hawking  and  G.F.R.  Ellis,  "The  Large   Scale
Structure   of  Space-time", Cambridge: Cambridge Univ. Press, 1973.}
\item{4.}{S.W. Hawking, {\it Phys. Rev.} {\bf D 37} (1988) 904;
S. Coleman, {\it   Nucl. Phys.} {\bf B 307} (1988) 864; {\it ibid.}
{\bf B 310} (1988) 643; S.B. Giddings and A. Strominger, {\it Nucl.
Phys.} {\bf B 307} (1988) 854.}
\item{5.} {C.G.~Callen, S.B.~Giddings, J.A.~Harvey and A.~Strominger,
{\it Phys. Rev.} {\bf D 45} (1992) R1005.}
\item{6.}V.P. Frolov and I.G. Novikov, {\it Phys. Rev.} {\bf D 42}
(1990) 1057.
\item{7.}   {T. Dray and G. 't Hooft, {\it Nucl. Phys.} {\bf B 253}
(1985) 173; T. Dray and G. 't Hooft, {\it Commun. Math. Phys.}
{\bf 99} (1985) 613; G.'t Hooft, {\it Nucl. Phys.} {\bf B 335} (1990) 138.}
\item{8.} {C.R. Stephens, G. 't Hooft and B.F. Whiting, "Black hole
evaporation without information loss", Utrecht/Gainesville prepr.
THU-93/20; UF-RAP-93-11; gr-qc/9310006.}
\item{9.} {G.~'t Hooft, "On the Quantization of Space and Time",
Proc. of  the  4th Seminar on Quantum Gravity, May 25-29, 1987,
Moscow, USSR, ed. M.A.~Markov et al (World Scientific 1988)}.
\item{10.} G. 't Hooft, {\it Nucl.Phys.} {\bf B 256} (1985) 727.
\item{11.}G. 't Hooft, {\it Physica Scripta} {\bf T 36} (1991) 247.
\item{12.}{G. 't  Hooft, ``A Two-dimensional
Model  with Discrete General Coordinate-Invar\-
iance",  in  ``The  Gardener of  Eden",  Physicalia Magazine, vol
{\bf 12}, in honour of R. Brout, eds. P. Nicoletopoulos and J. Orloff,
Brussels, 1990.}
\item{13.} G. 't Hooft, K. Isler and S. Kalitzin, {\it Nucl. Phys.}
{\bf B 386} (1992) 495

\end